\documentclass{ifacconf}
\usepackage{mathtools}
\usepackage{graphicx}      % include this line if your document contains figures
\usepackage{natbib}        % required for bibliography
\usepackage{algorithm}
\usepackage{algpseudocode}
\usepackage{multirow}
\usepackage{amsmath}
\usepackage{svg}
\usepackage{pdfpages}
\usepackage{subfig}
\usepackage{gensymb}
\usepackage{lipsum}% http://ctan.org/pkg/lipsum

\DeclareMathOperator*{\argmin}{arg\,min}

\begin{document}
\begin{frontmatter}

\title{Optimal control barrier functions for RL based safe powertrain control} 

%\thanks[footnoteinfo]{Sponsor and financial support acknowledgment
%goes here. Paper titles should be written in uppercase and lowercase
%letters, not all uppercase.}

\author[First]{Habtamu Hailemichael} 
\author[First]{Beshah Ayalew} 
%\author[First]{Lindsey Kerbel}
\author[Second]{Andrej Ivanco}
%\author[Second]{Keith Loiselle}

\address[First]{Automotive Engineering, Clemson University, Greenville, SC 29607, USA (hhailem, beshah)@clemson.edu.}
\address[Second]{Allison Transmission Inc., One Allison Way, Indianapolis, IN, 46222, USA (andrej.ivanco)@allisontransmission.com}

\begin{abstract}                % Abstract of not more than 250 words.

Reinforcement learning (RL) can improve control performance by seeking to learn optimal control policies in the end-use environment for vehicles and other systems. To accomplish this, RL algorithms need to sufficiently explore the state and action spaces. This presents inherent safety risks, and applying RL on safety-critical systems like vehicle powertrain control requires safety enforcement approaches. In this paper, we seek control-barrier function (CBF)-based safety certificates that demarcate safe regions where the RL agent could optimize the control performance. In particular, we derive optimal high-order CBFs that avoid conservatism while ensuring safety for a vehicle in traffic. We demonstrate the workings of the high-order CBF with an RL agent which uses a deep actor-critic architecture to learn to optimize fuel economy and other driver accommodation metrics. We find that the optimized high-order CBF allows the RL-based powertrain control agent to achieve higher total rewards without any crashes in training and evaluation while achieving better accommodation of driver demands compared to previously proposed exponential barrier function filters and model-based baseline controllers.

\end{abstract}


\end{frontmatter}
%===============================================================================

Conventional control design is dominated by rule-based and model-based approaches. Often intensive calibration is needed to accommodate diverse use environments of the controlled system. It is also common to design a control system to achieve certain standards, which again results in a system that performs well in the vicinity of standard design conditions \citep{Berry2010, Fontaras2017}. To alleviate these problems, efforts to adapt control policies to the end-user environment have garnered elevated interest following advances in logging quality data, such as with connected vehicle technology. Reinforcement learning (RL) is one of those methods that can capitalize on this deluge of data to learn and customize control policies. In addition to control policy adaptation, RL uses expressive learning models to capture complex environments and relationships, which are often poorly captured in traditional model-based approaches \citep{Brunke2021}.

For a given vehicle, powertrain control performance is affected by vocation, geography, loading, and traffic variabilities. Recent proposals on RL-based powertrain control focusing on adaptive cruise control \citep{Jurj2021} and eco-driving \citep{Wegener2021} have shown that RL can achieve a near-optimal solution by enhancing fuel economy, driver accommodation, and providing adaptation to the changing vehicle operating conditions. RL explores the state-action space systematically to find optimum control patterns and strategies that result in a higher reward. Such explorations, however, are often unbounded with respect to safety constraints and physical bounds. This limits RL's usage in practical safety-critical applications like powertrain control. 

Methods of instilling safety into RL training could be broadly categorized as hierarchical and integral approaches. Integral approaches introduce a safety cost (risk index) in each state-action and directly consider safety into the RL objective optimization \citep{Altman, Chow2015}. On the contrary, hierarchical methods allow the task-oriented controller to be optimized independently in a predefined safe-set and utilize a separate safety controller (or filters) to filter out unsafe actions proposed by the task-oriented controller. Safety filters must predict the proposed actions' safety consequences before designating a state-action pair safe or unsafe. Such prediction is not an easy task as it entails considerations of single-step safety or safety over the horizon, the effect of combinations of actions, uncertainties, time delay, and other considerations.

One could use external knowledge derived from offline data or the dynamic model of the system to design safety filters so that they can make sound safety predictions. Of the model-based approaches, CBFs provides scalability and minimal computation \citep{li2021comparison}. We have demonstrated successful integration of RL and CBFs for powertrain control in our previous works on ACC \citep{Hailemichael2022b} and driver assistance systems \citep{Hailemichael2022}. CBF guarantees safety by making the controller work in the invariant safe-set defined by a superlevel set of a continuously differentiable function $h(x) : R^n \to R$. The actions selected by the task/performance-oriented RL controllers are projected into the safe set in such a manner that the proposed actions are modified minimally, and no unsafe actions are passed to the controlled system.  Designing a CBF for powertrain control should  consider the fact that the system is of high relative degree by explicitly considering inertial and related  effects. Such CBFs are mostly handcrafted, but we could also use more systematic approaches that are tailored for high relative degree systems, as in the case of exponential CBFs \citep{Nguyen2016} and high order CBFs (HOCBFs) \citep{Xiao2021}.

 In general, the the safe state-action space may be too complex to capture in its entirety by a CBF of a certain structure. To maximize the space within which to the task-oriented controller can operate and optimize performance, CBFs should be designed to capture the largest portion of the safe set. In the case of powertrain control, the task-oriented RL controller aims to optimize fuel economy by selecting optimal gears and traction torque while accommodating driver demand and comfort. These objectives are encoded as reward terms that are then scalarized into a single value, by weighting them according to their importance. If not designed well, CBFs could limit the fulfillment of these objectives by conservatively limiting fast approach rates and high accelerations despite those being within the possible safe region of operation. As compared to ECBF, high-order CBFs provide a general equation structure, and more opportunity to expand the safe region. In this work, we derive a HOCBF for vehicle powertrain control and optimize the parameters of the HOCBF to maximize the possible safe set captured by the CBF without inducing undue conservatism. Optimizing the HOCBF entails selecting proper functions and parameters that, by filtering out unsafe actions, enable safe operations under extreme conditions (extreme possible load and vehicle parameters) considering the actuation bounds. Due to the widening of the safe region, the total reward  (so that vehicle performance) is shown to be enhanced as compared to non-optimized versions (such as ECBF).
 
The rest of the paper is organized as follows. Section \ref{sec: II} describes our derivation of the HOCBF as a safety filter for powertrain control and optimization of the safe set captured by the CBF. Section \ref{sec: III} describes the architecture of an RL agent combined with the HOCBF filter (RL-HOCBF) as a powertrain controller. Section \ref{sec: IV} discusses the implementation results of our RL-HOCBF arrangement on a medium-duty truck, and Section \ref{sec: V} concludes the paper.

\section{High-order CBF Safety Filter for powertrain control} \label{sec: II}
Consider a nonlinear control affine system:
\begin{equation} \label{eq:affinedynamics}
{{\dot{x}=f\left(x\right)+g\left(x\right)u,}}
\end{equation}
where $f$ and $g$ are locally Lipschitz, $ x\in\mathcal{R}^n $ is the system state, $ u\in\mathcal{R}^m $ is the control inputs. Assume a safe set defined by $\mathcal{C}=\left\{x\in\mathcal{R}^n|h\left(x\right)\geq0\right\}$, where $h:\mathcal{R}^n\rightarrow\mathcal{R}$ is a continuously differentiable function. Then $h$ is a control barrier function (CBF) if there exists an extended class $ {\mathcal{K}}_\infty$ function $\alpha$ such that for all $x\in Int\left(\mathcal{C}\right)=\left\{x\in\mathcal{R}^n:h\left(x\right)>0\right\}:$
\begin{equation} \label{eq:CBF}
{{
\displaystyle\sup_{u\in U}{\left[L_fh\left(x\right)+L_gh\left(x\right)u\right]}\geq-\alpha\left(h\left(x\right)\right)
}}.
\end{equation}

Considering the longitudinal motion of the vehicle, the collision avoidance problem (in a car following type scenario) is modeled with the state variables of separation distance $z$, the velocity of the host vehicle $v_h$ and velocity of the leading vehicle $v_l$, i.e. $x = \{z,v_h, v_l\}$. The state equations are: 
\begin{subequations}\label{eq:ACCsetup}
  \begin{equation}
    \label{eq:ACCsetup-a}
      {{\dot{z}=v_l-v_h},}
  \end{equation}
  \begin{equation}
    \label{eq:ACCsetup-b}
    {{\dot{v_l}=a_{l}},}
  \end{equation}
  \begin{equation}
    \label{eq:ACCsetup-c}
    {{{\dot{v}}_h=\frac{T_t}{r_wm_v}-\frac{F_r\left(v_h,m_v,\theta\right)}{m_v}},}
  \end{equation}
\end{subequations}

\begin{equation} \label{eq:resistanceforces}
{{
F_r=\frac{\rho A c_dv_h^2}{2}+m_vgf\cos{\theta}+m_vg\sin{\theta}
,}}
\end{equation}

where $F_r$ is the total resistance force including gravitational, rolling and aerodynamic resistances, and $T_t$ is the traction torque at the wheels. The parameters $c_d$, $f$, $\theta$, $m_v$ , $\rho$, $A_v$, $r_w$, $a_{l}$ are aerodynamic coefficient, rolling resistance coefficient, road grade, mass of the vehicle, density of air, frontal area of the vehicle, radius of the wheels, and acceleration of the leading vehicle, respectively. The above model could be readily put in the control affine form (\ref{eq:affinedynamics}) with $x=[z,v_l,v_h]'$.

The above definition of CBF is straightforward to apply to a system of relative degree one. For higher relative degree systems, such as the model of vehicle powertrain (which is of relative degree 2), we need high-order CBFs to take into account effects such as inertia. For a twice differentiable function $h: \mathrm{R}^n\to \mathrm{R}$, we can define a family of functions ${v}_i : \mathrm{R}^n\to \mathrm{R}, i\in \{1,2\}$ and associated sets ${\mathcal{C}_i}$ as
\begin{subequations}\label{eq:HOMPO}
  \begin{equation}
    \label{eq:HOMPO-a}
      {v_0(x)= h(x), \quad \mathcal{C}_0 = \{x\in \mathrm{R}^n:v_0(x)\ge 0\},}
  \end{equation}
  \begin{equation}
    \label{eq:HOMPO-b}
      {v_1(x)= \dot{v}_{0}+\alpha_1(v_{0}), \quad  \mathcal{C}_1= \{x\in \mathrm{R}^n:v_1(x)\ge 0\},}
  \end{equation}
  \begin{equation}
    \label{eq:HOMPO-c}
    {v_2(x)= \dot{v}_{1}+\alpha_2(v_{1}),  \quad \mathcal{C}_2= \{x\in \mathrm{R}^n:v_2(x)\ge 0\},}
  \end{equation}
\end{subequations}

where $\alpha_i$ is a class $\mathcal{K}$ function (a strictly increasing Lipschitz continuous function with $\alpha_i(0) = 0$). If $\mathcal{C}_2$ is forward invariant, then $\mathcal{C}_1$ is forward invariant when $x_0\in \mathcal{C}_1\cap\mathcal{C}_2$. The same relationship holds between $\mathcal{C}_1$ and $\mathcal{C}_0$. This leads to the conclusion that if $\mathcal{C}_2$ is forward invariant and $ x_0\in \mathcal{C}_0\cap \mathcal{C}_1\cap \mathcal{C}_2$, then $\mathcal{C}_0$ is forward invariant \citep{Xiao2021, Ames2019}. If we define $h(x)\geq 0$ as the safety constraint of the system, the fact that $\mathcal{C}_2$ is forward invariant prevents the system from leaving the safe set $\mathcal{C}_0$. In the present application, for the safety constraint $h(x)=z-z_0 \ge 0$, the analogous recursive functions $v_i$ and sets $\mathcal{C}_i$ are given by: 
\begin{subequations}\label{eq:HOMPO}
  \begin{equation}
    \label{eq:v0}
       {v_0(x)= z-z_0, \quad  \mathcal{C}_0= \{x\in \mathrm{R}^n:v_0(x)\ge 0\},}
  \end{equation}
  \begin{multline}\label{eq:v1}
      v_1(x)= v_l - v_h + {\alpha}_1(v_{0}), \\ 
      \mathcal{C}_1 = \{x\in \mathrm{R}^n:v_1(x)\ge 0\},
  \end{multline}
  \begin{multline}
    \label{eq:v2}
     v_2(x)= a_{l} + \frac{F_r\left(v_h,m_v,\theta\right)}{m_v} - \frac{T_t}{m_vr_w} + \dot{v}_{1} +\alpha_2(v_{1}),\\
     \mathcal{C}_2= \{x\in \mathrm{R}^n:v_2(x)\ge 0\}.
  \end{multline}
\end{subequations}

The task of HOCBF design is then to find functions $\alpha_1$ and $\alpha_2$ to make the $\mathcal{C}_2$ forward invariant within the action limit of the system. In addition, since the system should initialize and operate within $ x_0 \in \mathcal{C}_0\cap\mathcal{C}_1\cap \mathcal{C}_2$, the design of $\alpha$ functions should also consider maximizing the safe set captured by the intersection. Maximizing the operational area of the system under the application of HOCBF requires maximizing each set defined by the recursive functions. Note that $\mathcal{C}_0$ is the definition of safety without considering system dynamics and action limits, and the $z_0$ is chosen according to the designer's preference considering the safety and traffic efficiency implications. To design $\mathcal{C}_1$ and $\mathcal{C}_2$, we can use the relationships between the limiting relative velocity and relative distance, as we show below.

\begin{figure}
\includegraphics[width=7cm, height=5.25cm]{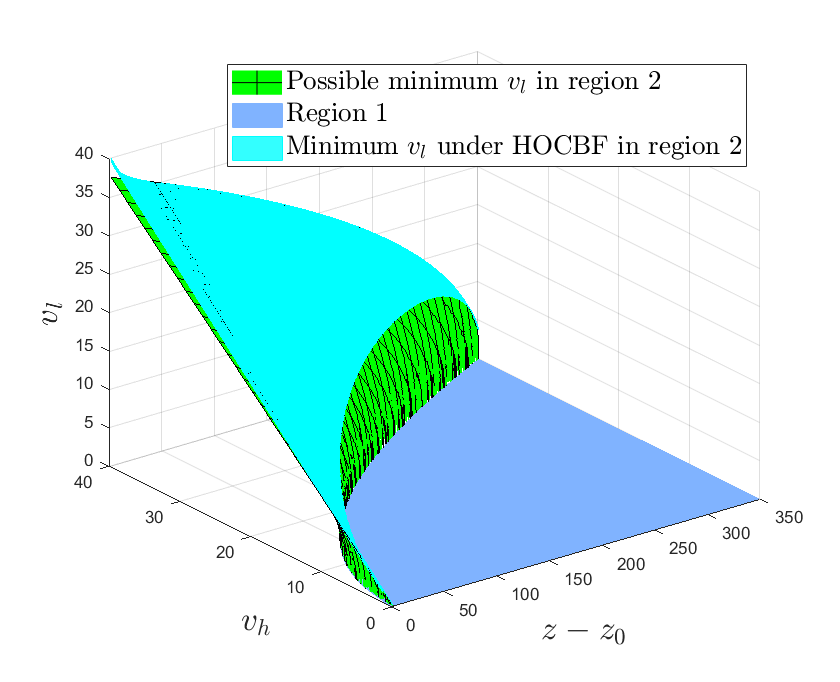}  
%\vspace*{-4mm}
\caption{Possible safe operating points against forward collision in a car-following scenario are divided into two regions (blue in region 1 and hatched green in region 2, see text).} 
\label{fig:Possiblesafesetvehicle}
\end{figure}

We can compute the possible safe set in which the host vehicle could avoid forward collision by considering the case that it applies maximum braking while the preceding vehicle is under its maximum braking. For each combination of separation distance and host vehicle velocity, Fig.\ref{fig:Possiblesafesetvehicle} shows the minimum velocity profile of the preceding vehicle necessary to avoid collision. For host vehicle velocities with relatively large separation distance, the preceding vehicle could be of zero velocity, and the host could use its maximum braking effort to stop the vehicle before collision. This region (region 1, shown in blue) is defined by (\ref{eq:region1}) where the stopping distance is less than the given separation. However, when the separation is below the stopping distance of the host vehicle's velocity, the required preceding vehicle velocity starts to increase ($v_l \ge 0$) to avoid collision (region 2, shown in green). Considering no reverse driving, region 2 is given by (\ref{eq:region2}). 
  \begin{equation}
    \label{eq:region1}
       {v_h^2 \leq -2a_{h,max} (z-z_0)},
  \end{equation}
  \begin{equation}
    \label{eq:region2}
      {v_l^2 \geq \frac{a_{l,max}}{a_{h,max}} v_h^2  + 2a_{l,max} (z-z_0)},
  \end{equation}

where $a_{h,max}$ and $a_{l,max}$ are the maximum deceleration of the host and the leading vehicle, respectively. For this analysis, the maximum deceleration is considered to be due to the maximum braking torque while the vehicle is at its maximum loading in a downhill drive.
  \begin{equation}
    \label{eq:ahd}
      {a_{h,max} = \frac{T_{b-max}}{r_vm_{v-max}} + m_{v-max}gsin\left( \theta_{max} \right)}.
  \end{equation}

Maximizing $\mathcal{C}_1$ requires maximizing the function $v_1$  by finding a bound on the relative velocity ($v_l - v_h$) for a given relative distance ($v_0$). For region 1, the extreme relative velocity is when the preceding vehicle is stopped and the limiting relative velocity ($v_l - v_h$) is given by
  \begin{equation}
    \label{eq:v1_region1}
       {v_{rel-lim,1}=v_h=-\sqrt{-2a_{h,max}\ (z-z_0)}}.
  \end{equation}

For region 2, the relative velocity for a given host vehicle velocity and relative distance can be shown to be:
  \begin{equation}
    \label{eq:vrel2}
      {v_{rel-lim,2}=\sqrt{\frac{a_{l,max}}{a_{h,max}}v_h^2 + 2a_{l,max}(z-z_0)}-v_h}.
  \end{equation}
We can further find the host vehicle velocity with the limiting relative velocity for a given separation distance as (\ref{eq:vrel2_lim}), and use this to find the relationship between the relative distance and the extreme relative velocity profile in region 2 (\ref{eq:vrel_lim_min}).
    \begin{equation}
    \label{eq:vrel2_lim}
      {v_{h-lim}=\min(v_{h-max},\sqrt{\frac{2a_{h,max}\ (z-z_0\ )}{a_{l,max}/a_{h,max}\ -1}})}.
  \end{equation}
    \begin{equation}
    \label{eq:vrel_lim_min}
      {v_{rel-min,2}=\sqrt{\frac{a_{l,max}}{a_{h,max}}v_{h-lim}^2 + 2a_{l,max}(z-z_0)}-v_{h-lim}}.
  \end{equation}

   % \begin{multline}
   % \label{eq:v1_region2}
    %  {v_{rel-min,2}=-\sqrt{\frac{4a_{l,max}^3a_{h,max}\left(z-z_0\right)^2}{\left(a_{l,max}-1\right)^2} - 2a_{l,max}\left(z-z_0\right)} \\
     % +\sqrt{\frac{2a_{l,max}(z-z_0)}{\frac{a_{l,max}}{a_{h,max}}-1}}}
  %\end{multline}

\begin{figure}
\begin{center}
\includegraphics[width=6cm, height=4 cm]{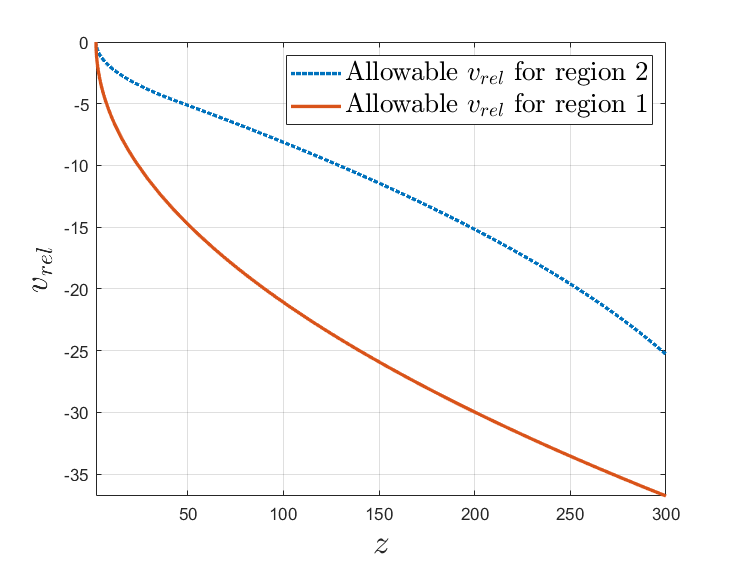}    
%\vspace*{-3mm}
\caption{The allowable relative velocities for a given relative distance for region 1 and region 2.} 
\label{fig:Relativevandd}
\end{center}
\end{figure}

The limiting relative velocity vs. relative distance profiles ($z$) for a particular medium-duty truck is shown in Fig.\ref{fig:Relativevandd}. It is possible to see that region 2 has a lower (in magnitude) allowable relative velocity for a given separation distance. Accordingly, we can define the function $v_1$ to capture the maximum $\mathcal{C}_1$ in terms of the relative distance and velocity:
\begin{equation}
    \label{eq:v1_region2_}
       {v_1 = v_l-v_h + \alpha_1(z -z_0) =  v_l-v_h - v_{rel-min,2}}.
\end{equation}
The design of $v_2$ considers maximization of $\mathcal{C}_2$ and making sure it is forward invariant within the action limits of the system. Having optimized $\mathcal{C}_1$, it is sufficient to make sure $\mathcal{C}_1 \subseteq \mathcal{C}_2$ to maximize $ \mathcal{C}_0\cap\mathcal{C}_1\cap\mathcal{C}_2$ and a simple choice is $\mathcal{C}_1 = \mathcal{C}_2$. Noticing $\alpha_2$ is zero at the boundary of $\mathcal{C}_1$ (since $v_1 = 0$) and the maximum time derivative of $v_{rel-min,2}$ is $a_{l,max} -a_{h,max}$, (\ref{eq:v2}) transforms to
\begin{equation}
\begin{aligned}
\label{eq:v2zero}
v_2 = & \, a_l + \frac{F_r(v_h, m_v, \theta)}{m_v} - \frac{T_t}{m_v r_w} \\
& - (a_{l,\text{max}} - a_{h,\text{max}}) \geq 0.
\end{aligned}
\end{equation}

at the boundary. As this relation could be satisfied within the system action limits, the set $\mathcal{C}_2$ is forward invariant and gives the freedom to use any choice of $\alpha_2$.

The operational region demarcated by $\mathcal{C}_0\cap\mathcal{C}_1\cap\mathcal{C}_2$ is the maximum possible set under the structure of HOCBF that guarantees safety under the worst-case scenario. The conservativeness of the design is pronounced as the maximum deceleration of the leading vehicle $(a_{l,max})$ increases, resulting in more and more possible safe operation points fail to be captured by the HOCBF, especially in region 1. To mitigate this conservativeness, we propose a separate HOCBF for each region. HOCBF definition of $v_1$ in (\ref{eq:v1_region2_}) is dedicated for region 2. For region 1, we can use the stopping distance relationship in (\ref{eq:v1_region1}) to determine the allowable relative velocity, and define $v_1$ as
  \begin{equation}
    \label{eq:v1_region1_}
      {v_1=v_l - v_h + \sqrt{-2a_{h,max}\ (z-z_0)}}
  \end{equation}
  
Such separation makes the HOCBF in region 1 more permissive to allow fast approaches and, as a result, gives more opportunity to accommodate the driver's desired accelerations. Notice that the boundary between region 1 and region 2 is safe with respect to each region's definition of HOCBF filter. Using the same reasoning that led to (\ref{eq:v2zero}), $\mathcal{C}_2$ of region 1 is forward invariant for a region captured by $\mathcal{C}_1$.

Finally, our HOCBF based safety filter enforces safety by projecting the traction torque action proposed by the RL agent $T_a\left(s\right)$ (see next section) to the closest traction torque $T_t$ that is consistent with the definition of $\mathcal{C}_2$ ($v_2\left(x\right)\geq 0$). This is done by posing and solving the quadratic program summarized as:
\begin{equation}
\label{eq:optimization}
\begin{aligned}
 T^{*}_t = & \displaystyle\argmin_{T_{t}} \frac{1}{2}\left \| T_{t}-T_{a}(s) \right \|^2 \\
\textrm{s.t.} \quad & v_2 = a_{l} + \frac{ F_r\left(v_h,m_v,\theta\right)}{m_v} - \frac{T_t}{m_vr_w}+ \dot{v_1} +\alpha_2\left(v_1\right)\geq0
\end{aligned}
\end{equation}
For region 1 where $v_h\le\sqrt{-2a_{h,max}\ (z-z_0\ \ )}$
  \begin{equation}
    \label{eq:HOCBF_region1}
      {v_1=v_l-v_h+\sqrt{-2a_{h,max}(z-z_0\ )}}.
  \end{equation}
For region 2 where $v_h > \sqrt{-2a_{h,max}\ (z-z_0\ \ )}$
  \begin{align}
    \label{eq:HOCBF_region2}
      {v_1=v_l-v_h\ - v_{rel-min,2}}.      
  \end{align}
\section{Vehicle Environment and RL agent } \label{sec: III}

To formulate the RL agent for powertrain control, the system is modeled as Markov decision process (MDP) consisting of states $s$, actions $a$, a reward function $r\left(s,a\right)$, and discounting factor $\gamma$ \citep{Sutton2018}. The vehicle environment is modeled by states that include the ego vehicle velocity, the relative velocity between the preceding and ego vehicle, the driver demanded acceleration, the actual vehicle acceleration, the separation distance with the preceding vehicle, transmission gear, mass of the vehicle, road grade, previous torque applied and a flag to show if the vehicle is in radar range $f$, i.e., $s=\{v_l, v_{rel}, a_{des}, a, z, n_{g}, m_v, \theta, T_p, f\}$ \citep{Hailemichael2022}. We use an actor-critic architecture in which the actor (parameterized by ${\boldsymbol\theta}$) determines the control policy for a given state $\pi\left(s\middle|{\boldsymbol\theta}\right)$ and a critic (parameterized by ${\boldsymbol\phi}$) evaluates these actions by providing the associated action values $Q\left(s,a\middle|{\boldsymbol\phi}\right)$. The actor-network (the RL agent) makes decision on the traction torque at the wheel $T_a$ and gear change $\Delta n_g $, i.e. $a=\{T_a,\Delta{n_g}\}$, based on the state inputs. For the gear change, the RL agent outputs three discrete outputs for the available gear changes \{upshift, staying in gear, downshift\} and categorical sampling is then used to obtain the gear change policy (\ref{eq:gearcatagory}). On the other hand, the RL agent outputs the mean and variance for the continuous traction torque, which is sampled from a normal distribution (\ref{eq:torqeguassian}). 
\begin{equation} \label{eq:gearcatagory}
{
\pi_{\boldsymbol\theta}^g(\Delta n_g|s)=Cat(\alpha_{\boldsymbol\theta}(s)),\forall s\ \sum_{k=1}^{3}{\alpha_{k,{\boldsymbol\theta}}\left(s\right)=1}.
}
\end{equation}
\begin{equation} \label{eq:torqeguassian}
{
\pi^{T}_{\boldsymbol\theta\left(T_a|s\right)}=\mathcal{N}\left(\mu_{\boldsymbol\theta}\left(s\right),\sigma_{\boldsymbol\theta}^2\left(s\right)\right) 
}.
\end{equation}

\begin{figure}
\begin{center}
\includegraphics[width=8cm, height=5cm]{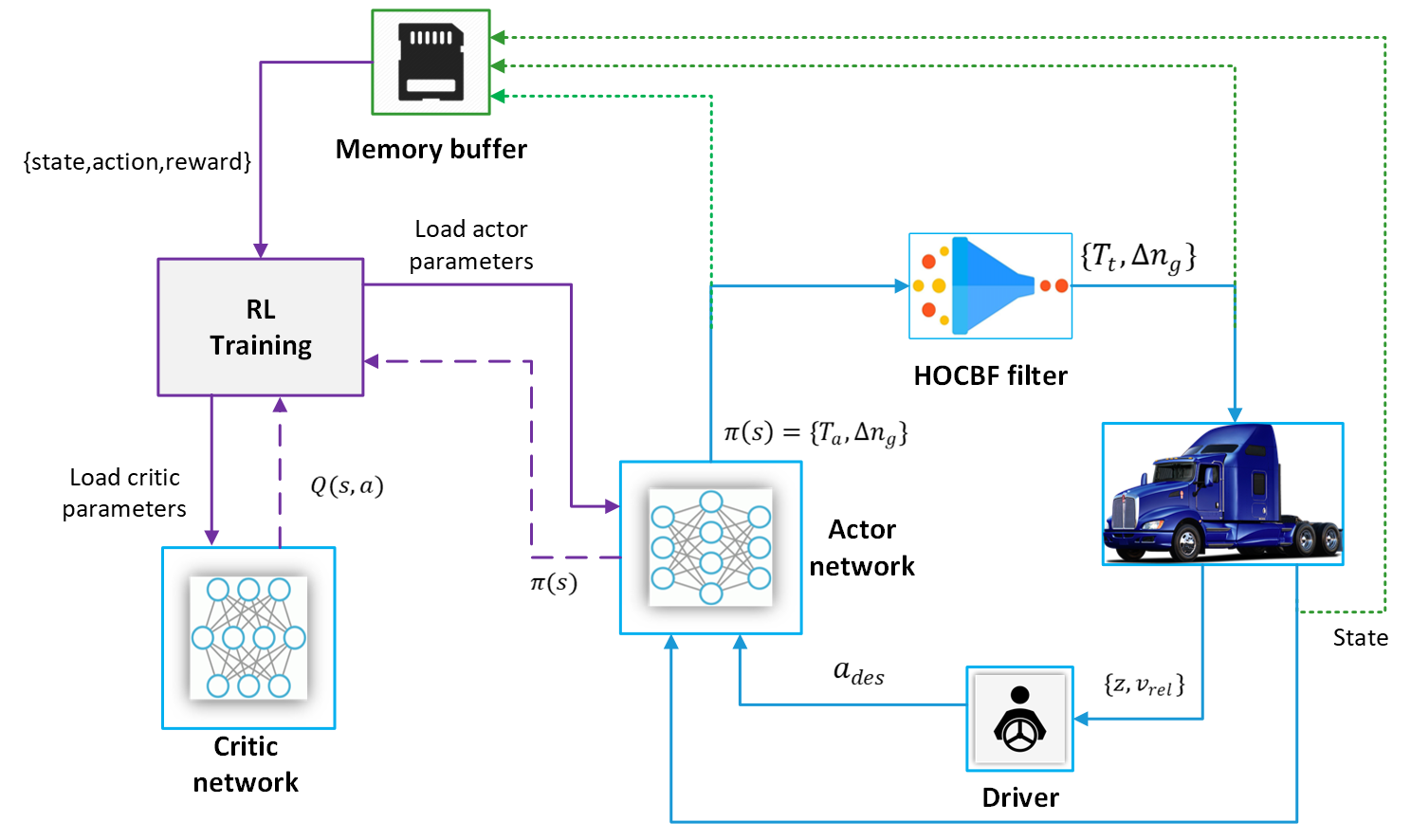}    
\caption{Schematic of the RL agent and its training} 
\label{fig:Training}
\end{center}
\end{figure}

To ensure only safe traction torque is sent to the vehicle environment for implementation, the HOCBF safety layer projects the potentially unsafe RL action proposals $T_a$ to safe traction torque demand $T_t$ (\ref{eq:optimization}) to be sent to the vehicle's powertrain, as shown by Fig.\ref{fig:Training}. The engine torque and engine speed that brings about this wheel traction torque are then calculated utilizing transmission ratios of the selected gear, the final drive and drive-line efficiency, and the associated fuel rate logged (read from the fuel map in our simulations).

 The reward for the RL agent is designed to accommodate the driver's acceleration requests in a fuel-efficient and smooth operation. These different objectives are captured by dedicated reward terms weighted according to their importance as:
\begin{equation} \label{eq:outofrange}
\begin{split}
r = & w_a 0.1^{\frac{|a - a_{\text{des}}|}{a_{\text{des,max}}}} + w_{f} 0.1^{\frac{\dot{m}_f}{\dot{m}_{f,\text{max}}}} \\
& + w_{T} 0.1^{\frac{|\Delta T_t|}{\Delta T_{t,\text{max}}}} + w_{g} 0.1^{|\Delta n_{g}|},
\end{split}
\end{equation}

where $ \dot{m}_f $ and $\Delta{T_{t}}$ are the fuel rate and torque change, respectively. The driver accommodation term ensures fulfillment of the driver's acceleration request, weighted by a relatively higher $w_a$. The fuel rate reward term, weighted by $w_{f}$, encourages the RL agent to operate in fuel-efficient operating points. Gear hunting and the associated rough vehicle operation are mitigated by including a gear shifting penalty term weighted by $w_g$. Oscillatory responses are discouraged by including a term that penalizes abrupt changes of traction torque, weighted by $w_t$. Note that the relevant reward signals are normalized by their corresponding maximum values, as noted by the max subscripts.

State, action and reward data are continuously stored in a memory buffer and later used in trainings. Since the RL agent outputs continuous traction torque and discrete gear change actions, the training algorithm needs to effectively accommodate hybrid action space. Even if it is possible to use other state-of-the-art algorithms like Proximal Policy Optimization (PPO) \citep{Schulman2017}, we found Hybrid Maximum A Posteriori Policy Optimization (HMPO) is found to be a good fit \citep{Kerbel2022,Neunert2020}. In addition to being scalable and robust, the fact that it is off-policy makes it data efficient to apply to real-world RL trainings. We refer the reader to \cite{Neunert2020}for implementation details of MPO and  our paper \cite{Kerbel2022} for its raw application in powertrain control.

%\begin{equation} \label{MPOtwo}
%\theta_{k+1} = \arg\max_{\theta} \mathbb{E}_{s	\sim R} \left[ KL \left(q(a|s) || %\pi_{\theta}(a|s) \right) \right] \\
%\end{equation}. 

\section{Results and Discussions } \label{sec: IV}
\begin{figure*}%
  \centering
    \subfloat[\centering RL training progress]{{\includegraphics[width=6.5cm, height=5cm]{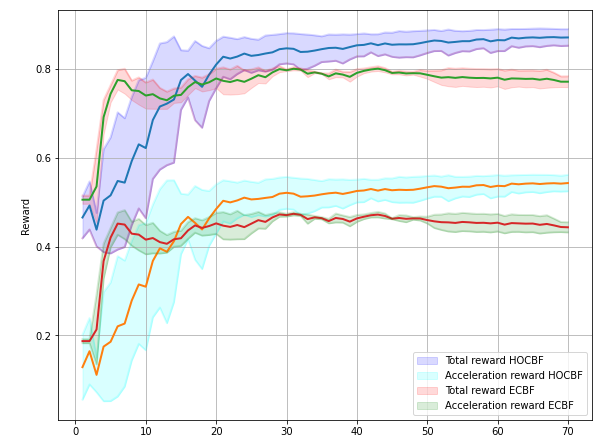} }}%
    \subfloat[\centering HOCBF enhances driver accomodation]{{\includegraphics[width=5.0cm, height=5cm]{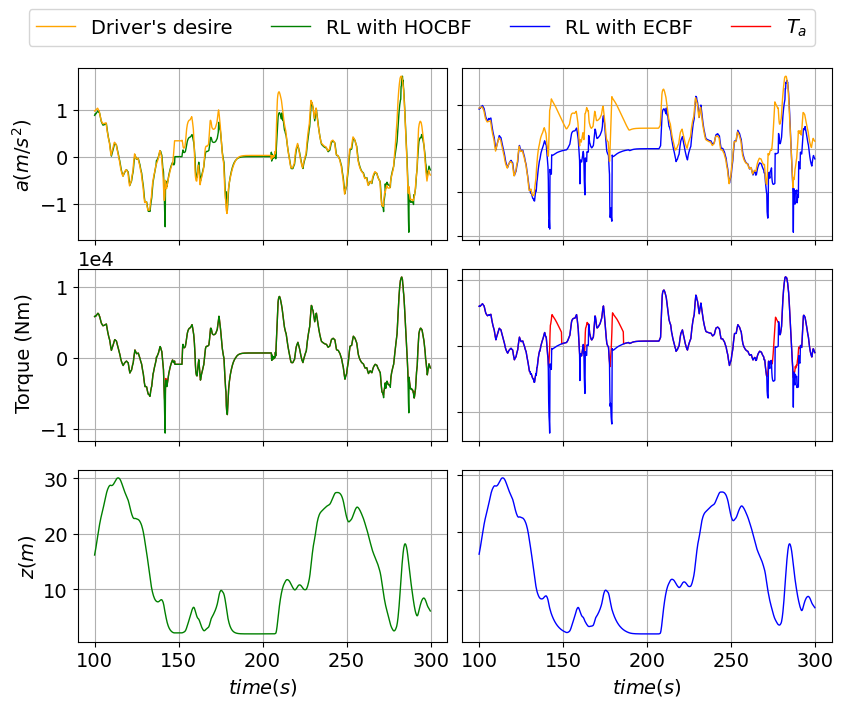} }}%
        \subfloat[\centering Evaluation of RL-HOCBF]{{\includegraphics[width=6cm, height=5cm]{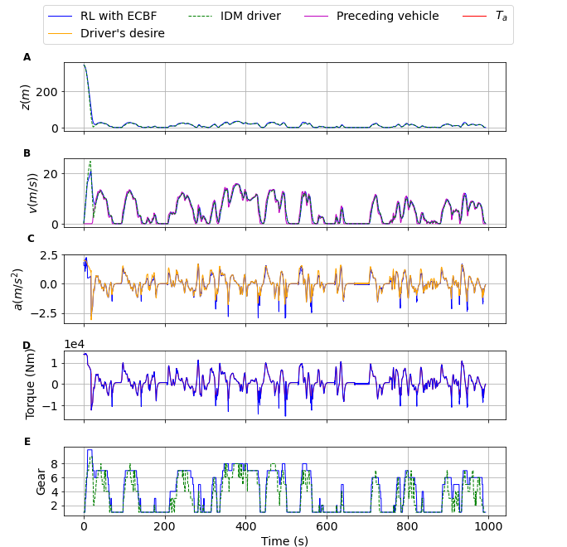} }}%
    \caption{ \centering Training of  RL agent with HOCBF filter and performance comparison.}%
\label{fig:results} 
\end{figure*}
The optimized HOCBF described above is applied with an RL-based powertrain controller on a model of medium-duty truck. The truck has a 10-speed automated manual transmission (AMT) with a weight range of 5 to 12 tons. Using the braking limit of $15000 Nm$, and wheel radius of $0.5 m$, the maximum deceleration $a_{h,max}$ of the truck is calculated using (\ref{eq:ahd}) to be $2.27 m/s^2$ in a flat road. Using maximum host velocity $v_{h-max}$ of $40 m/s$ and the maximum deceleration of preceding vehicle $a_{l,max}$ $2 m/s^2$, the possible operational area is shown by the Fig.\ref{fig:Possiblesafesetvehicle}. Based on (\ref{eq:v1_region1}) and (\ref{eq:vrel2}), the limiting relative velocities for a given separation distance are generated, as depicted in Fig.\ref{fig:Relativevandd}, and used to optimize $\mathcal{C}_1$. It is possible to see that the allowable relative velocity profile at region 2 is more restrictive, and hence, a separate HOCBF is considered to ameliorate conservatives as already discussed. Using a linear function for $\alpha_2$, we obtain:
\begin{equation}
\label{eq:HOCBFv1}
\begin{aligned}
& For \quad v_h \leq \sqrt{-4.54(z-z_0)} \\
& \quad v_1 = v_l - v_h + \sqrt{-4.54(z-z_0)} \\
& For \quad v_h > \sqrt{-4.54(z-z_0)} \\
& \quad v_1 = v_l-v_h\ + min\left( 40,\sqrt{33.6 (z-z_0)} \right) - \\ 
& \sqrt{0.88\bullet min\left( 40,\sqrt{33.6 (z-z_0)} \right)^2 + 4.54(z-z_0)} 
\end{aligned}
\end{equation}
\begin{equation}
\label{eq:v2final}
\begin{aligned}
{v_2 = a_{l,max} +\frac{F_r\left(v_h,m_v,\theta\right)}{m_v} -\frac{T_t}{m_vr_w}+\dot{v}_1+2v_1}
\end{aligned}
\end{equation}
The cascaded connection of the RL task controller along with HOCBF safety filter (RL-HOCBF) is trained in a car-following scenario. The driver of the host vehicle is modeled by the intelligent driver model (IDM) following a preceding vehicle under the FTP drive cycle. In each simulation step, as shown in Fig.\ref{fig:Training}, the IDM requests acceleration based on the current relative velocity and distance. Seeing the environment's state and the IDM driver's desire, the actor network proposes traction torque and gear change actions. As mentioned before, the HOCBF filters out the unsafe traction torques by projecting unsafe actions to the closest safe action using the quadratic programming problem in (\ref{eq:optimization}). The safe traction torque, along with the gear change selections, are then implemented in the simulated vehicle environment, and the associated rewards are observed. The different objectives are encoded into the rewards using weights of $[w_a=0.675,w_f=0.25,w_T=0.075,w_g=0.075]$ for the results we present below.

 The HMPO algorithm's hyper parameter settings and the vehicle model parameters are given in Table.\ref{tb:parametersMPO}. In order to help capture different driving experiences, the training data is randomized by adding noise to the velocity profile of the preceding vehicle, using randomized initial separation distance, road grade and mass, and manipulating the parameters of the driver model. 

HOCBF ensures no unsafe actions are sent to the vehicle environment, and as a result, the vehicle never comes closer than the safe separation distance ($z_0=2$).
Fig.\ref{fig:results}a shows the progress of multiple RL trainings with an optimized HOCBF filer in comparison with trainings with an exponential CBF (ECBF) safety filter, which results when we use linear $\alpha_1$ and $\alpha_2$ functions in the HOCBF filter as in (\ref{eq:ECBF}) (The ECBF design for this application has been extensively discussed in our previous paper \cite{Hailemichael2022}). The optimized HOCBF gives more room for the RL agent to fulfill the acceleration requests of the driver and results in a higher driver accommodation reward. This fact is again demonstrated by Fig.\ref{fig:results}b, in which the optimized HOCBF allows faster approaches to the preceding vehicle, whereas its ECBF counterpart projects the traction torques to decrease the approach rate.
\begin{equation}
\begin{aligned}
\label{eq:ECBF}
& v_2 = a_{l} + \frac{F_r(v_h,m_v,\theta)}{m_v} - \frac{T_t}{m_vr_w} + k_1(z - z_0) + k_2(v_l - v_h)
\end{aligned}
\end{equation}

The performance of the well-trained RL-HOCBF controller is compared with a model-based baseline in Table \ref{tb:Performance}, in which the preceding vehicle follows ArtUrban drive cycle (different from the one used in training). For the baseline powertrain controller, we formulate a simple feedforward controller where the torque required to compensate for resistances and the torque to provide the requested acceleration are summed to give the traction torque input to the environment. Unlike the RL controller, which has no knowledge of the engine's fuel consumption map or any of the modeled dynamics, the baseline controller selects an optimal gear with the lowest fuel rate with the full knowledge of the engine fuel consumption map according to a scheme described in \cite{Yoon2020}. The comparison of the root-mean-square error between the driver's desire and the actual acceleration shows that the RL-HOCBF controller attains better driver accommodation. The RL agent also learns fuel-efficient operating points and manages to improve the MPG by $7.6\%$ in our evaluation setup. Fig.\ref{fig:results}c shows that the RL agent demonstrates eco-driving attributes by learning to frequent higher gear operations and implement early shiftings.
\begin{table}[hb]
\begin{center}
\caption{Vehicle environment and RL hyperparameter setting}\label{tb:parametersMPO}
\begin{tabular}{cc|cc}
\hline
\multicolumn{2}{| c |}{Vehicle Parameters} & \multicolumn{2}{| c |}{MPO Hyperparameters}  \\\hline
Mass & 5\ - 12 tons & Actor, critic learning rate & ${10}^{-4},{10}^{-5}$ \\
$A_u$ & $7.71m^2$ & Dual constraint& 0.1 \\
$C_d$& $0.08$ & Retrace steps& 1 \\
$r_w$& $0.5$ & KL constraints $\epsilon_\mu,\epsilon_\sigma,\epsilon_d$ & $0.1,0.001,0.1$ \\
$f$& $0.015$ & $\alpha_d,\alpha_c$ & 10\\
$a_{l}$&$-2$ & $\gamma$ & 0.95 \\\hline
\end{tabular}
\end{center}
\end{table}
\begin{table}[hb]
\begin{center}
\caption{Performance comparison between model-based baseline and RL-based controller with HOCBF safety filter.}
\label{tb:Performance}
\begin{tabular}{p{1.25cm}|p{3cm}p{3cm}}
\hline
& Baseline & IDM with RL-HOCBF \\\hline
MPG & {\quad}$ 6.875 (-)$ &{\quad}{\quad} $7.4 (+7.6\%)$ \\
$a_{rms}$ & {\quad}{\quad} $0.38$ &{\quad}{\quad} $0.2$ \\ \hline 

\end{tabular}
\end{center}
\end{table}
\section{Conclusion} \label{sec: V}
In this paper, we derived high-order control barrier functions (HOCBF) for powertrain control and maximized the safe set captured by the HOCBF, considering worst-case scenarios akin to reachability analysis. The proposed traction torque action of the RL agent is projected to the closest safe traction torque, and no unsafe actions are permitted to pass to the controlled environment. Implementation on a simulated medium-duty truck demonstrated that when RL based powertrain controller is combined with optimized HOCBF, it allows fast approaches that help to enhance driver accommodation compared to simplified exponential CBFs derived for the same system. In addition, a comparison with a model-based powertrain control baseline showed that the RL-HOCBF arrangement significantly enhanced fuel economy and driver accommodation.

Future works seeks to extend the optimal HOCBF design framework to more degrees of freedom in the vehicle motion (including steering and full automation).

%\subsection{Copyright Form}

%IFAC will put in place an electronic copyright transfer system in due
%course. Please \emph{do not} send copyright forms by mail or fax. More
%information on this will be made available on IFAC website.

%\begin{ack}
%Place acknowledgments here.
%\end{ack}

\bibliography{bibliography}             % bib file to produce the bibliography
                                                     % with bibtex (preferred)
                                                   
%\begin{thebibliography}{xx}  % you can also add the bibliography by hand

%\bibitem[Able(1956)]{Abl:56}
%B.C. Able.
%\newblock Nucleic acid content of microscope.
%\newblock \emph{Nature}, 135:\penalty0 7--9, 1956.

%\bibitem[Able et~al.(1954)Able, Tagg, and Rush]{AbTaRu:54}
%B.C. Able, R.A. Tagg, and M.~Rush.
%\newblock Enzyme-catalyzed cellular transanimations.
%\newblock In A.F. Round, editor, \emph{Advances in Enzymology}, volume~2, pages
%  125--247. Academic Press, New York, 3rd edition, 1954.

%\bibitem[Keohane(1958)]{Keo:58}
%\newblock \emph{Power and Interdependence: World Politics in Transitions}.
%\newblock Little, Brown \& Co., Boston, 1958.

%\bibitem[Powers(1985)]{Pow:85}
%\newblock Is there a way out?
%\newblock \emph{Harpers}, pages 35--47, June 1985.

%\bibitem[Soukhanov(1992)]{Heritage:92}
%\newblock \emph{{The American Heritage. Dictionary of the American Language}}.
%\newblock Houghton Mifflin Company, 1992.

%\end{thebibliography}

%\appendix
%\section{A summary of Latin grammar}    % Each appendix must have a short title.
%\section{Some Latin vocabulary}              % Sections and subsections are supported  
                                                                         % in the appendices.
\end{document}